\documentclass[twocolumn,prb,showpacs,floatfix,superscriptaddress]{revtex4}
\usepackage{graphicx}
\usepackage{bm}
\usepackage{amsmath,amssymb}

\newcommand{\be}{\begin{equation}}
\newcommand{\ee}{\end{equation}}
\newcommand{\bea}{\begin{eqnarray}}
\newcommand{\eea}{\end{eqnarray}}

\begin{document}

\title{Excitation spectra and ground state properties of the layered spin-$1/2$ frustrated antiferromagnets
Cs$_2$CuCl$_4$ and Cs$_2$CuBr$_4$}

\author{John~O.~Fj{\ae}restad}
\affiliation{Department of Physics, University of Queensland,
Brisbane, Qld 4072, Australia}
\author{Weihong Zheng}
\affiliation{School of Physics, University of New South Wales,
Sydney NSW 2052, Australia}
\author{Rajiv R.~P.~Singh}
\affiliation{Department of Physics, University of California, Davis,
CA 95616, USA}
\author{Ross H. McKenzie}
\affiliation{Department of Physics, University of Queensland,
Brisbane, Qld 4072, Australia}
\author{Radu Coldea}
\affiliation{Department of Physics, University of Bristol, Bristol
BS8 1TL, United Kingdom}

\date{\today}

\pacs{75.10.Jm}

\begin{abstract}
We use series expansion methods to study ground- and excited-state
properties in the helically ordered phase of spin-$1/2$ frustrated
antiferromagnets on an anisotropic triangular lattice. We calculate
the ground state energy, ordering wavevector, sublattice
magnetization and one-magnon excitation spectrum for parameters
relevant to Cs$_2$CuCl$_4$ and Cs$_2$CuBr$_4$. Both materials are
modeled in terms of a Heisenberg model with spatially anisotropic
exchange constants; for Cs$_2$CuCl$_4$ we also take into account the
additional Dzyaloshinskii-Moriya (DM) interaction. We compare our
results for Cs$_2$CuCl$_4$ with unpolarized neutron scattering
experiments and find good agreement. In particular, the large
quantum renormalizations of the one-magnon dispersion are well
accounted for in our analysis, and inclusion of the DM interaction
brings the theoretical predictions for the ordering wavevector and
the magnon dispersion closer to the experimental results.
\end{abstract}

\maketitle

\section{Introduction}
\label{intro}

The study of frustrated quantum antiferromagnets is central to
modern condensed matter physics. Much of the interest in these
many-body systems stems from the possibility, first envisaged by
Anderson,\cite{andfaz} that their quantum fluctuations may be so
strong as to lead to exotic ``spin-liquid'' ground states
characterized by the absence of broken symmetries of any kind.
Another hallmark signature of spin liquids is the existence of
deconfined fractionalized spin-$1/2$ excitations, usually referred
to as spinons. Unfortunately it has been very hard to find
experimental realizations of such states. Important progress has
however been made recently with the identification of some materials
which may have, or at least be \textit{close} to having, a
spin-liquid ground state. Two such promising candidates are the
organic compound
$\kappa$-(BEDT-TTF)$_2$Cu$_2$(CN)$_3$\cite{shimizu} and the
layered $S=1/2$ frustrated antiferromagnet Cs$_2$CuCl$_4$.

The magnetic properties of Cs$_2$CuCl$_4$ have been extensively
studied using neutron scattering.\cite{coldea} At low temperatures
$T<T_N=0.62$ K long-range helical magnetic order is observed and the
low-energy excitation spectra contain relatively sharp modes
characteristic of magnons, the expected Goldstone modes. The magnon
dispersion does however show very strong renormalizations compared
to linear spin-wave theory. Furthermore, the dominant feature of the
spectra is in fact a broad continuum, occurring at medium to high
energies, which carries most of the spectral weight and which
persists also for temperatures above $T_N$. The interpretation of
this continuum has recently been the subject of much
debate.\cite{coldea,chung1,bocquet,zhouwen,chung2,isakov,veillette,dalidovich,alicea,yunsor,sb}

The most interesting hypothesis, already suggested in Ref.
\onlinecite{coldea}, is that the continuum is due to two-spinon
scattering resulting from Cs$_2$CuCl$_4$ being close to a quantum
phase transition to a two-dimensional spin-liquid state. A number of
theoretical proposals have been made regarding the nature of the
spin liquid that might be involved in such an unconventional
scenario.\cite{chung1,zhouwen,chung2,isakov,alicea,yunsor} An
alternative hypothesis, that the effects of magnon-magnon scattering
included within a standard (nonlinear) spin-wave approach might be
able to explain the experimental results, was recently explored in
Refs. \onlinecite{veillette} and \onlinecite{dalidovich}. While
qualitative features of the calculated spectra were found to be
similar to the experimental ones, the quantitative agreement was
however not satisfactory. In particular, the obtained quantum
renormalization of the magnon dispersion (calculated to lowest order
in magnon-magnon interactions) was significantly lower than found
experimentally.

\begin{figure}[h]
\begin{center}
  \includegraphics[width=6cm]{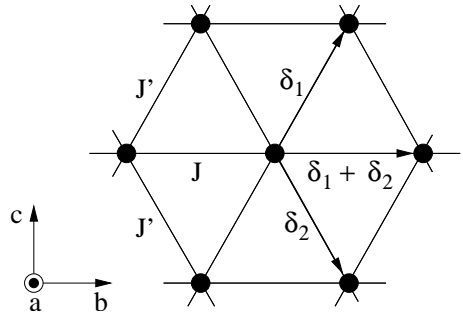}
  \caption{\label{trlattice} Exchange constants
  $J$, $J'$ and lattice basis vectors $\bm{\delta}_1$, $\bm{\delta}_2$
  on a triangular lattice. The standard notation for axes in
  Cs$_2$CuCl$_4$ is also indicated.}
\end{center}
\end{figure}

The magnetic interactions in Cs$_2$CuCl$_4$ are sufficiently weak
that the external magnetic field required to fully polarize the
spins is experimentally accessible. By measuring the magnon
dispersion relation in this fully-polarized state (in which quantum
fluctuations are completely suppressed by the applied field), it was
found\cite{coldea-prl-02} that the layers in Cs$_2$CuCl$_4$ are
nearly decoupled from each other and well described by a spin-$1/2$
triangular-lattice Heisenberg antiferromagnet with exchange
constants $J$ and $J'$ (defined in Fig. \ref{trlattice}) with
$J'/J\approx 1/3$, weakly perturbed by an additional intra-layer
interaction of Dzyaloshinskii-Moriya (DM) type,\cite{dm} of strength
$D/J\approx 0.05$. Although weak, the DM interaction has several
important consequences, one being that the ordering plane of the
spins coincides with the plane of the triangular lattice because the
DM interaction makes the latter an``easy" plane.

Theoretical studies of Cs$_2$CuCl$_4$ and models relevant to it have
been carried out using many different methods. For the zero-field
case, which is our focus here, these methods include spin-wave
theory,\cite{merino,veillette,dalidovich} series expansions,\cite{zheng99}
variational approaches,\cite{yunoki,yunsor} and various field-theoretical
approaches.\cite{chung1,bocquet,zhouwen,chung2,isakov,alicea,sb,ceccatto}
While most of these studies only considered the Heisenberg part of
the Hamiltonian, some of them also investigated the effects of the
DM interaction. Refs. \onlinecite{veillette} and
\onlinecite{dalidovich}, using the nonlinear spin-wave approach,
found that several quantities, including the ordering wavevector and
the quantum renormalization of the magnon dispersion, were quite
sensitive to the DM interaction, and that the reduced spin rotation
symmetry of the Hamiltonian for $D\neq 0$ also leads to a gap in the
magnon dispersion at the ordering wavevector. Ref.
\onlinecite{dalidovich} also studied the sublattice magnetization as
a function of $D$ (for values of $J$, $J'$ appropriate for
Cs$_2$CuCl$_4$) and found that it vanishes when $D/J<0.008$, and
hence that the DM interaction is crucial to stabilize the magnetic
order in Cs$_2$CuCl$_4$. The same qualitative conclusion was reached
in Ref. \onlinecite{sb} which used a quasi-one-dimensional approach,
treating the interchain interactions ($J'$ and $D$) using a
perturbative renormalization group analysis; these authors found
that in the absence of the DM interaction the ground state was
dimerized. Finally, the easy-plane nature of the Hamiltonian when
$D\neq 0$ plays an essential role in the algebraic vortex-liquid
scenario proposed for Cs$_2$CuCl$_4$ in Ref. \onlinecite{alicea};
also in this study the magnetic order was found to be driven by the
DM interaction.

Another interesting material is Cs$_2$CuBr$_4$, which can be
described in terms of a spin-$1/2$ triangular-lattice Heisenberg
antiferromagnet with $J'/J\approx 1/2$.\cite{zheng05hight} Compared
to Cs$_2$CuCl$_4$ this material is therefore more frustrated (i.e.,
closer to the isotropic limit $J=J'$) and further away from the
one-dimensional limit $J\gg J'$. The thermodynamic properties of
both materials, particularly the uniform magnetic susceptibility,
have been extensively studied. There are also a large number of
organic materials from superconducting families that have insulating
magnetic phases that can be described by a spatially anisotropic
triangular-lattice model.\cite{powell} Recent high temperature
series expansion studies of these systems\cite{zheng05hight} found
that a large class of the organic materials appeared close to the
isotropic triangular-lattice limit, while the inorganic materials
were closer to the one-dimensional limit.

In this paper we study spin-$1/2$ triangular-lattice
antiferromagnets using zero-temperature high-order series
expansions. We focus on the case $J > J'$ for which, according to
our previous series expansion study in Ref. \onlinecite{zheng99},
the ground state has noncollinear (helical) long-range magnetic
order and the ordering wave vector varies continuously with the
model parameters. We calculate the ground state energy, ordering
wavevector, sublattice magnetization, and magnon dispersion for
values of $J'/J$ relevant to Cs$_2$CuCl$_4$ and Cs$_2$CuBr$_4$. For
Cs$_2$CuCl$_4$ we also consider the effects of the DM interaction.
We compare our results with those of other theoretical approaches
and with neutron scattering experiments. A few of the results
discussed in this paper have already been briefly presented in Ref.
\onlinecite{zheng06}.

The outline of the paper is as follows. In Sec. \ref{model} we
describe the model Hamiltonian. In Sec. \ref{series} we discuss the
series expansion method used for studying zero-temperature
properties in the helical phase. In Secs. \ref{gse} and \ref{disp}
we discuss our results for ground state properties and the magnon
dispersion, respectively. Finally, our conclusions are presented in
Sec. \ref{concl}.

\section{Model Hamiltonian}
\label{model}

The antiferromagnetic spin-$1/2$ Heisenberg model with spatially
anisotropic exchange constants on a triangular lattice forms the
basis for a theoretical description of the magnetic properties of
both Cs$_2$CuCl$_4$ and Cs$_2$CuBr$_4$. The Heisenberg Hamiltonian
reads
\begin{equation}
H_{\text{H}} = \sum_{\bm{R}}[J\;\bm{S}_{\bm{R}}\cdot
\bm{S}_{\bm{R}+\bm{\delta}_1+\bm{\delta}_2}+J' \;
\bm{S}_{\bm{R}}\cdot
(\bm{S}_{\bm{R}+\bm{\delta}_1}+\bm{S}_{\bm{R}+\bm{\delta}_2})],
\label{HH}
\end{equation}
where $\bm{S}_{\bm{R}}$ is the spin operator on site $\bm{R}$, $J$
and $J'$ are the antiferromagnetic exchange constants, and
$\bm{\delta}_1$ and $\bm{\delta}_2$ are nearest-neighbor lattice
vectors, as defined in Fig. \ref{trlattice}.

When the ground state of $H_{\text{H}}$ has magnetic long-range
order, the spin expectation value can be written \be \langle
\bm{S}_{\bm{R}}\rangle =M[\bm{n}_1\cos (\bm{Q}\cdot \bm{R})+
\bm{n}_2 \sin (\bm{Q}\cdot \bm{R})] \label{helical} \ee where $M$ is
the sublattice magnetization, $\bm{n}_1$ and $\bm{n}_2$ are two
arbitrary orthogonal unit vectors, and $\bm{Q}=(0,Q,0)$ is the
ordering wavevector with $\cos (Q b/2)=\cos q$, where $q$ ($2q$)
is the angle between nearest-neighbor spins coupled by $J'$ ($J$)
and $b$ is the lattice constant in the $\hat{b}$ direction (see Fig.
\ref{trlattice}). For classical spins (corresponding to the limit of
spin $S\to\infty$), $M=S$ and $q=\arccos(-J'/2J)$ (for $J>J'/2$).
Eq. (\ref{helical}) then implies that the magnetic order is helical,
with the ordering plane spanned by the vectors $\bm{n}_1$ and
$\bm{n}_2$. For quantum spins, quantum fluctuations could in
principle kill the magnetic order. However, in one of our earlier
series expansion studies\cite{zheng99} of the model (\ref{HH}) for
$S=1/2$, the magnetic order appeared to be robust as long as $J/J'$
is not too large. The quantum fluctuations do however renormalize
$Q$ (and hence $q$) away from the classical value and reduce the
strength of the ordering so that the sublattice magnetization $M<S$.

Cs$_2$CuBr$_4$ can be described by the Hamiltonian (\ref{HH}) with
$J/J'\approx 2$.\cite{zheng05hight} In contrast, Cs$_2$CuCl$_4$ has
$J=0.374(5)$ meV and $J'=0.128(5)$ meV, giving a more anisotropic
ratio $J/J'\approx 2.92$.\cite{coldea-prl-02} Furthermore, the spin
Hamiltonian of Cs$_2$CuCl$_4$ also contains a Dzyaloshinskii-Moriya
(DM) interaction\cite{dm} $H_{\text{DM}}$ of the
form\cite{coldea-prl-02} \be H_{\text{DM}}= \sum_{\bm{R}}\bm{D}\cdot
\bm{S}_{\bm{R}}\times (\bm{S}_{\bm{R}+\bm{\delta}_1}
+\bm{S}_{\bm{R}+\bm{\delta}_2}), \ee where $\bm{D}$ lies along the
$a$-direction, perpendicular to the plane of the triangular lattice,
with $D=0.020(2)$ meV, giving $D/J'\approx 0.16$. Thus the DM
interaction is numerically a relatively small perturbation on the
dominant Heisenberg Hamiltonian $H_H$. Nevertheless, the DM
interaction has several notable consequences. It breaks the full
SU(2) spin rotation symmetry of the Heisenberg part down to $U(1)$
by making the plane of the triangular lattice an "easy plane" which
therefore becomes the ordering plane of the spins. Furthermore, as
the DM interaction can be seen to give rise to a linear coupling to
the unit vector $\bm{n}_1\times \bm{n}_2$ (which points
perpendicular to the ordering plane), it selects a unique direction
for this vector, corresponding to a specific chirality or handedness
of the spin order. In Cs$_2$CuCl$_4$ the direction of $\bm{D}$ in
fact alternates from layer to layer\cite{coldea-prl-02} and hence
the chirality alternates as well.

When discussing Cs$_2$CuCl$_4$ in this paper we will consider both
the Hamiltonian $H_{\text{H}}+H_{\text{DM}}$ with $D/J'=0.16$ as
well as the more simplified model defined by neglecting the DM
interaction, i.e., $D=0$. In Cs$_2$CuCl$_4$ there is also an
antiferromagnetic interlayer exchange interaction
$J''$,\cite{coldea-prl-02} but because this is very small
($J''=0.017(2)$ meV $\approx J/45$) and because its effectiveness in
coupling the layers is further reduced by the alternating chirality,
we choose not to include this interlayer interaction in our
analysis, thus focussing on a single layer.

\section{Series expansions in the helically ordered phase}
\label{series}

In this section we discuss some aspects of the series expansion
analysis of the helically ordered phase.

Following Ref. \onlinecite{zheng0608008}, we assume that the spins
order in the $xz$ plane (which would be the $bc$ plane in
Cs$_2$CuCl$_4$), with $\bm{D}$ pointing in the perpendicular
direction. We rotate all the spins so as to have a ferromagnetic
ground state, with the resulting Hamiltonian $H$:
\begin{equation}
H = H_1 + J H_2 + H_3
\end{equation}
where
\be H_1 = J \cos{(2q)} \sum_{\langle in \rangle} S^z_i S^z_n +
 [ J'\cos{(q)} -D \sin(q) ] \sum_{\langle ij \rangle} S^z_i S^z_j,
\label{eq_H1} \ee
\be H_2 = \sum_{\langle in \rangle} S_i^y S_n^y + \cos{(2q)} S_i^x
S_n^x + \sin{(2q)}(S_i^z S_n^x - S_i^x S_n^z), \label{eq_H2} \ee
\bea
H_3 &=& \sum_{\langle ij \rangle} J' S_i^y S_j^y + [ J' \cos{(q)} - D \sin(q) ] S_i^x S_j^x \nonumber \\
&& + [ J' \sin{(q)} + D \cos(q) ] (S_i^z S_j^x - S_i^x S_j^z).
\label{eq_H3}
\eea
where the sum $\langle in \rangle$ is over
nearest-neighbor sites connected by ``horizontal'' bonds in Fig.
\ref{trlattice} with exchange interaction $J$, and the sum $\langle ij \rangle$
 is over nearest-neighbor sites connected by ``diagonal'' bonds
 with exchange interaction $J'$.

Next, we introduce the Hamiltonian
\begin{equation}
H(\lambda)\equiv H_0+\lambda V \label{eq_final}
\end{equation}
where
\begin{equation}
H_0 = H_1 - t\sum_i (S_i^z-1/2), \label{H0}
\end{equation}
\begin{equation}
V = J H_2 + H_3 + t\sum_i (S_i^z-1/2). \label{V}
\end{equation}
The last term of strength $t$ in both $H_0$ and $V$ is a local field
term, which can be included to improve convergence. $H(\lambda=0)$
is a ferromagnetic Ising model with two degenerate ground states,
while $H(\lambda=1)$ is the model whose properties we are interested
in. We use linked-cluster methods to develop series expansion in
powers of $\lambda$ for ground state properties and the magnon
excitation spectra.

For $J\neq J'$, the lattice has $C_{2v}$ symmetry (the symmetry
operations are the identity, inversion, and reflections about the
$b$ and $c$ axes), and the series for the spin-triplet excitation
spectra has the following form:
\be
\omega (k_x, k_y)/J' =
\sum_{r=0}^{\infty} \lambda^r \sum_{m,n} c_{r,m,n} \cos( \frac{m}{2}
k_x) \cos(\frac{n \sqrt{3}}{2} k_y )
\label{eq_mk}
\ee
where $c_{r,m,n}$ are series coefficients, $m$ and $n$ are
integers, representing a hopping over distance $(m/2, n\sqrt{3}/2)$,
and $m+n$ is a even number. For this case, series for the spin-triplet
dispersion has been computed to order $\lambda^{8}$, and the
calculations involve a list of  25022 linked-clusters, up to 9
sites. The series coefficients $c_{r,m,n}$ for $J=2.92$, $J'=1$,
$q=1.64$, $D=0.16$, and $t=4$
 are given in Table
\ref{tab_mk}.

For more details we refer to Ref. \onlinecite{zheng0608008} where series expansions
for the ground state properties and magnon disperion of the spin-$1/2$ Heisenberg model
on the anisotropic triangular lattice are discussed at length.\cite{oitmaa}

\begin{figure}[!htb]
\begin{center}
\includegraphics[width=6cm]{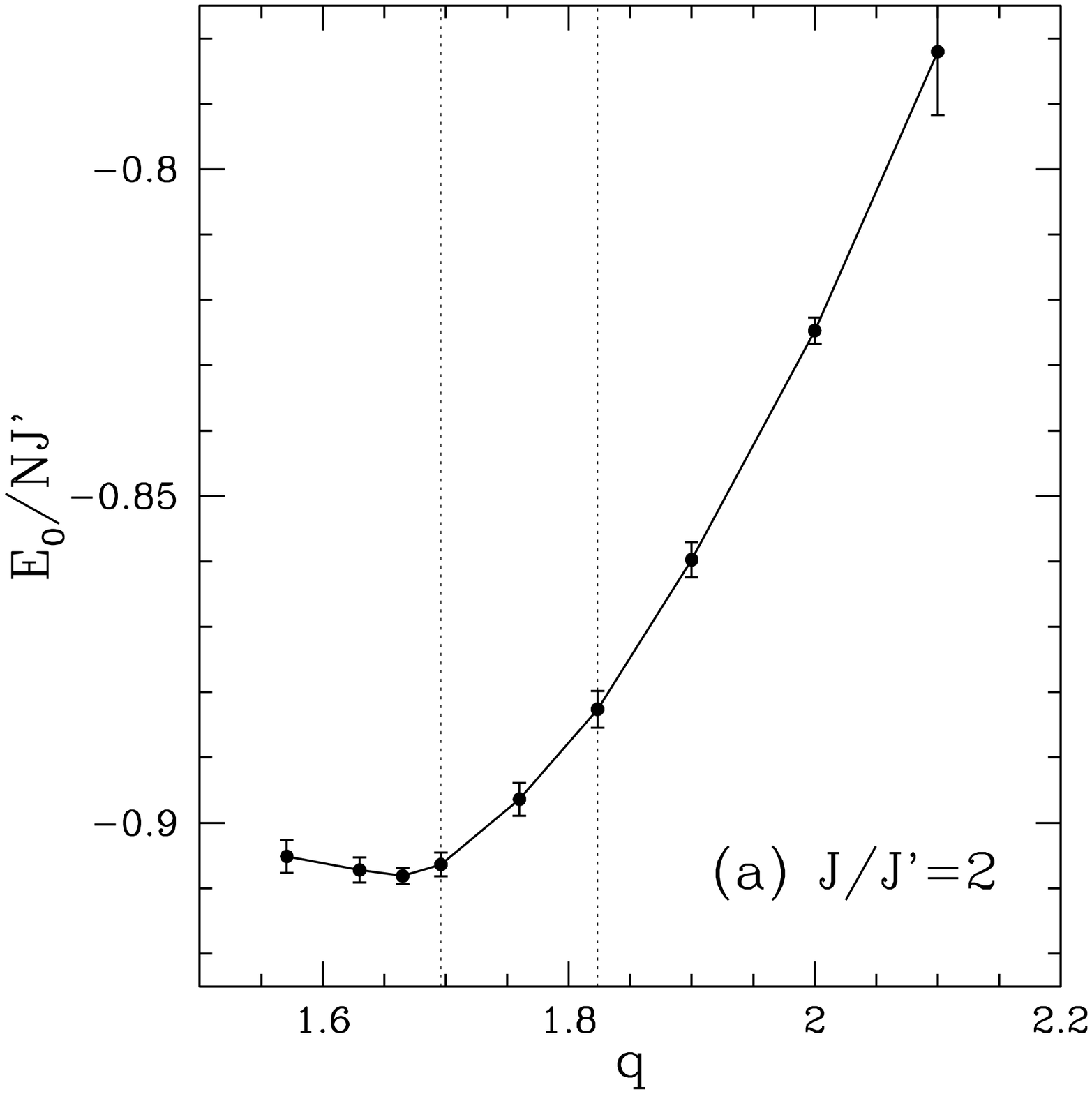}
\includegraphics[width=6cm]{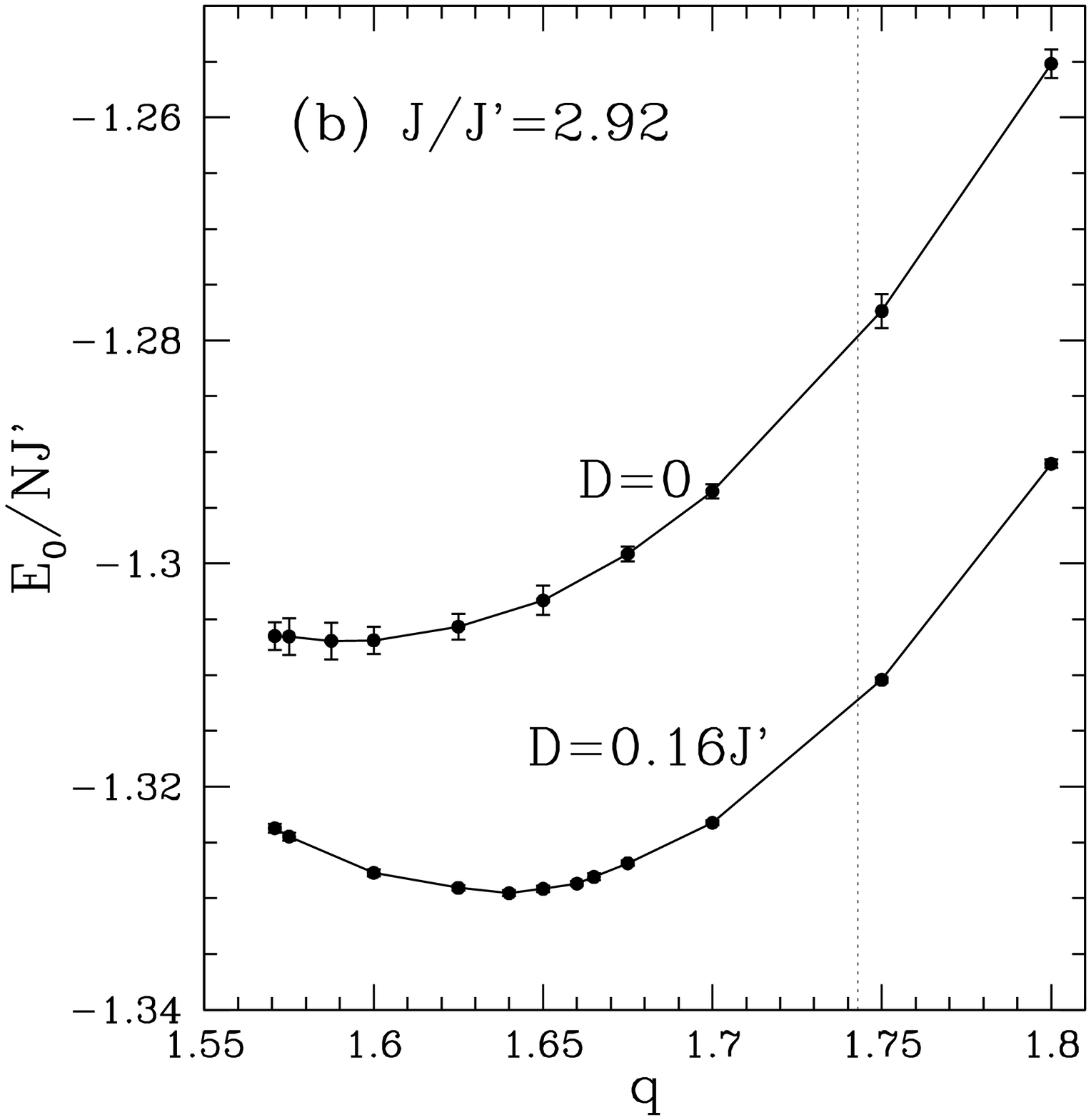}
\caption{\label{fig_e0} The ground state energy per site, as a
function of the angle $q$ between nearest-neighbor spins along $J'$
bonds, obtained from the series expansion around the Ising limit.
(a) $J/J'=2$, relevant for Cs$_2$CuBr$_4$. Here the left and right
vertical dashed lines denote the value of $q$ obtained from dimer
expansions [Eq. (16) in Ref. \onlinecite{zheng99}], and the
classical ground state, respectively. (b) $J/J'=2.92$, the ratio
appropriate for Cs$_2$CuCl$_4$. The DM interaction ($D=0.16J'$)
lowers the ground state energy and shifts the ordering angle from
$q\approx 1.59$ to $q\approx 1.64$.}
\end{center}
\end{figure}

\section{Ground state properties}
\label{gse}

\subsection{Ordering wavevector}

In Fig. \ref{fig_e0} we present results for the ground state energy
per site, as a function of the ordering angle $q$, for various model
parameters. The actual, realized, value of $q$ is that which
minimizes the ground state energy.

\textit{Cs$_2$CuBr$_4$} [Fig. \ref{fig_e0}(a)]. The value of $q$
predicted by the Ising expansion is seen to be considerably smaller
than the classical value (right dashed line) but is quite close to
the value predicted from a dimer expansion (left dashed line; Eq.
(16) in Ref. \onlinecite{zheng99}).

\textit{Cs$_2$CuCl$_4$} [Fig. \ref{fig_e0}(b)]. The ordering
wavevector $Q$ (equivalently, the ordering angle $q$) is conventionally
expressed in terms of the incommensuration $\epsilon$, defined as
the deviation between $Q$ and the antiferromagnetic (with respect
to the $b$ direction) wavevector $\pi/b$, measured in units of
$2\pi/b$: $\epsilon\equiv (Q-\pi/b)/(2\pi/b)= q/\pi-1/2$. For
$D=0$ the series calculation gives $q\approx 1.59$ which implies
$\epsilon\approx 0.006$, while for $D/J'=0.16$, $q\approx 1.64$
which gives $\epsilon\approx 0.022$. Thus inclusion of the DM
interaction increases the incommensuration and brings it much closer
to the experimental value $\epsilon=0.030(2)$.\cite{coldea-prl-02}

In Table \ref{values} we compare our results for the incommensuration
$\epsilon$ for the helically ordered phase in Cs$_2$CuCl$_4$ with
experiments and with predictions from some other theoretical
approaches.

\subsection{Sublattice magnetization}
\label{magnetization}

For Cs$_2$CuCl$_4$ ($J/J'=2.92$ and $D/J'=0.16$) we find the
sublattice magnetization to be $M=0.213(10)$. In this result the
main source of error is the uncertainty in the angle $q$. It is
interesting to compare this with predictions from other theoretical
approaches (see Table \ref{values}). In particular, by using the $1/S$
expansion and taking into account quantum corrections up to and
including order $1/S^2$ to the classical result, Ref. \onlinecite{dalidovich}
found the slighly smaller value $M\approx 0.18$ for the same parameters,
and also predicted that $M=0$ for $D=0$, i.e. that
quantum fluctuations would be strong enough to completely melt the
magnetically ordered state in the absence of the DM interaction. In
contrast, our analysis indicates a small but nonzero value $M\approx
0.10$ for the magnetization when $D=0$, although it should be noted
that the error bars are significant (see Fig. 6 in Ref.
\onlinecite{zheng99}). We conclude that in both approaches the DM
interaction is found to cause a considerable suppression of quantum
fluctuations and strengthen the magnetic ordering tendencies.
Including the interlayer coupling $J''$ is not expected to change
the magnitude of the ordered moment by much, as $J''$ is rather small
($J''/J\sim 45$) and also because the chirality of the spin ordering
is opposite in neighboring layers such that the interlayer coupling
energy in fact vanishes at the mean field level.

For Cs$_2$CuBr$_2$ ($J/J'=2$ and $D=0$) series expansions predict
$M\approx 0.17$.\cite{zheng99} It would be interesting to make
precise neutron
scattering measurements of the magnitude of the ordered spin moment
in both Cs$_2$CuCl$_4$ and Cs$_2$CuBr$_4$ to compare directly with
the theoretical calculations presented here.

\begin{table}[h]
\begin{center}
\caption{Comparison of values for the incommensuration $\epsilon$
and sublattice magnetization $M$ for Cs$_2$CuCl$_4$ obtained from experiments and various
theoretical approaches. Unless noted otherwise, the values of the
exchange interactions $J$ and $J'$ used in the theoretical
calculations are $J=0.374$ meV and $J'=0.128$ meV, corresponding to
a ratio $J/J'=2.92$. $D$ denotes the strength of the Dzyaloshinskii-Moriya interaction.} \label{values}
\begin{tabular}{|l|l|l|l|l|}\hline
 & \multicolumn{2}{c|}{$\epsilon$} & \multicolumn{2}{c|}{$M$} \\\cline{2-5}
 & $D=0.16J'$ & $D=0$ & $D=0.16J'$ & $D=0$ \\\hline\hline
Experiment & 0.030(2)\cite{coldea-prl-02} & N/A & & N/A \\
Classical & 0.0533 & 0.0547 & 0.5 & 0.5\\
$+1/S$ correction & 0.031\cite{veillette,dalidovich} & 0.022\cite{dalidovich}  &
$\approx 0.25$\cite{vcc,dalidovich} & $\approx 0.07$\cite{merino,dalidovich}\\
$+1/S^2$ correction & & 0.011\cite{dalidovich} & $\approx 0.18$\cite{dalidovich} & 0\cite{dalidovich} \\
Series & 0.022 & 0.006 & 0.213 & $\approx$ 0.1\\
Variational RVB\footnote{For a ratio $J'/J=0.33$.} & & 0.018\cite{yunoki} & &
0\cite{yunsor}\\\hline
\end{tabular}
\end{center}
\end{table}

\section{Magnon dispersion}
\label{disp}

\subsection{Cs$_2$CuCl$_4$}

In this subsection we present series expansion results for the
magnon dispersion for parameters relevant to Cs$_2$CuCl$_4$, discuss
the effects of the DM interaction on this dispersion, and compare it
to the dispersion obtained from spin-wave theory with $1/S$
corrections\cite{veillette,dalidovich} and to the experimental
dispersion obtained from inelastic neutron scattering with
unpolarized neutrons.\cite{coldea}

\subsubsection{Series dispersion}

\begin{figure}[!htb]
\begin{center}
  \includegraphics[width=7cm,bbllx=0,bblly=0,bburx=526,
  bbury=528,angle=0,clip=]{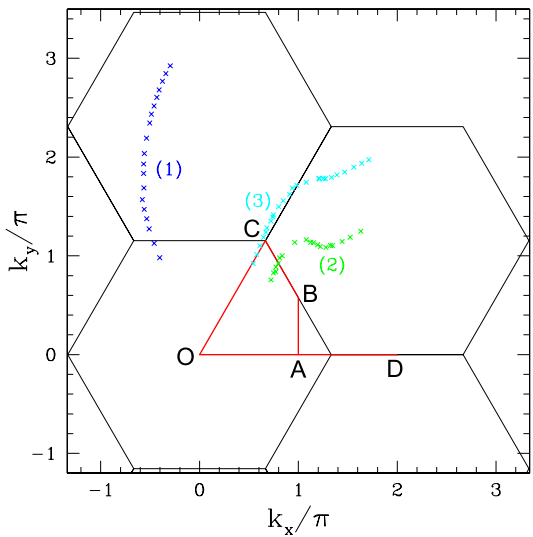}
  \caption{\label{trlattbz}
  (Color online) Reciprocal space diagram for the triangular
  lattice. The path ABCOAD ($A=(\pi,0)$, $B=(\pi,\pi/\sqrt{3})$,
$C=(2\pi/3,2\pi/\sqrt{3})$, $O=(0,0)$, and $D=(2\pi,0)$) and the
three sets of points denoted (1)-(3) are cuts along which magnon
dispersions are plotted in subsequent figures in the paper.}
\end{center}
\end{figure}

\begin{figure}[!htb]
\begin{center}
  \includegraphics[width=7cm,bbllx=20,bblly=151,bburx=554,
  bbury=706,angle=0,clip=]{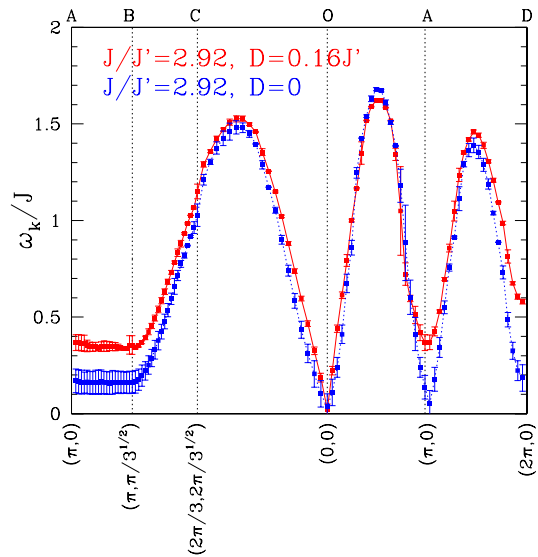}
  \caption{\label{fig_mk_y3}
(Color online) Magnon dispersion $\omega_{\bm{k}}$, as calculated
from series expansions, for parameters relevant to Cs$_2$CuCl$_4$,
plotted along the path ABCOAD shown in Fig. \ref{trlattbz}. When the
DM interaction is included the magnon energy is significantly
enhanced along AB and near the point D, and a gap opens up at the
ordering wavevector $\bm{k}=\bm{Q}$.}
\end{center}
\end{figure}

In Fig. \ref{fig_mk_y3} we show series expansion results for the
magnon dispersion along the $\bm{k}$-space path ABCOAD in Fig.
\ref{trlattbz}. The dashed blue curve is for $J/J'=2.92$ and $D=0$,
while the full red curve includes the effect of a finite $D/J'=0.16$.
We note the following features:

(i) Along AB, which is perpendicular to the chain direction ($b$
axis), the excitation energy for $D/J'=0.16$ is significantly
enhanced (by more than a factor of two) over that for $D=0$. The
dispersion along AB is very flat in both cases.

(ii) The $D\neq 0$ dispersion has a gap at $\bm{k}=\bm{Q}$ while
the $D=0$ dispersion is gapless there. This follows from symmetry
arguments. For $D=0$ the model has full SU(2) spin rotation
symmetry. The helical order then leads to gapless excitations
(Goldstone modes) at $\bm{k}=0$ and at the ordering vector
$\bm{k}=\bm{Q}$ . The Goldstone mode at $\bm{k}=\bm{Q}$ is
associated with rotations of the ordering plane of the spins, which
is arbitrary when $D=0$. The SU(2) symmetry is broken by the the DM
interaction which fixes the ordering plane to coincide with the
plane of the triangular lattice. Thus rotations of the ordering
plane costs a finite energy for $D\neq 0$, which creates a gap in
the magnon dispersion at
$\bm{k}=\bm{Q}$.\cite{nagamiya,rastelli,vcc,veillette}
(It is important to note, however, that to
get gapless excitations at the appropriate $\bm{k}$ vectors in the
series calculations we need to bias the analysis as discussed in
some detail in Ref. \onlinecite{zheng0608008}; an unbiased analysis
always gives a gap.)

(iii) Near the point D the excitation energy for $D/J'=0.16$ is
significantly enhanced (by approximately a factor of three) over the
$D=0$ case.

(iv) Overall, the high-energy parts of the dispersion are quite
insensitive to the DM interaction, while the low-energy parts are
quite sensitive. A notable exception to the latter is the region
around $\bm{k}=0$, since the Goldstone theorem dictates gapless
excitations at $\bm{k}=0$ regardless of whether $D$ is zero or not
(this is because the $\bm{k}=0$ Goldstone mode is associated with
long-wavelength rotations of the spins \textit{within} their
ordering plane\cite{nagamiya,rastelli,veillette,vcc}).

\begin{figure}[!htb]
\begin{center}
  \includegraphics[width=8cm]{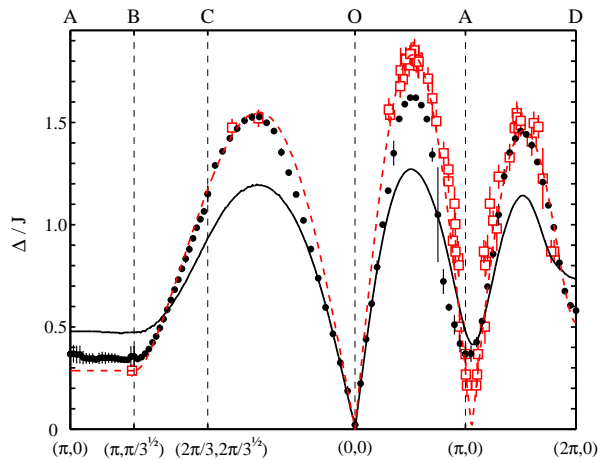}
  \caption{\label{fig-th-exp}
(Color online) Magnon dispersion for $J/J'=2.92$ and $D/J'=0.16$
(solid points from series, black solid line from spin-wave theory
with $1/S$ correction) compared to experimental dispersion in
Cs$_2$CuCl$_4$ (squares from Ref. \onlinecite{coldea}, dashed line
is experimental parameterization) along the path ABCOAD in Fig.
\ref{trlattbz}. Compared to the dispersion from spin-wave theory,
the series dispersion is enhanced along the $J$ bonds and decreased
perpendicular to them. Note that theoretical and experimental
dispersions cannot be directly compared near the ordering wavevector
$\bm{k}=\bm{Q}$, as the gap in $\omega_{\bm{k}}$ expected at this
wavevector cannot be resolved in the unpolarized neutron scattering
experiments (see text about how the experimental dispersion is
extracted).}
\end{center}
\end{figure}

\subsubsection{Comparison with spin-wave theory}

Next we compare our theoretical dispersion for $D/J'=0.16$ with that
obtained from spin-wave theory with $1/S$ corrections (LSWT + $1/S$
for short).\cite{veillette,dalidovich} Both dispersions are plotted
in Fig. \ref{fig-th-exp} (black dots and full line, respectively).
We see that compared to the spin-wave prediction, the excitation
energy is increased in the $b$ direction along which
neighboring spins are coupled by the strong $J$ bonds, and decreased
in the perpendicular $c$ direction. This corresponds to an
upwards renormalization of $J$ and downwards renormalization of $J'$
with respect to the spin-wave prediction, which thus effectively
makes the system appear more one-dimensional. (The renormalizations
of $J$ and $J'$ with respect to \textit{linear} spin-wave theory
(LSWT) are even bigger,\cite{zheng06} as the LSWT + $1/S$ dispersion
itself is in turn characterized by an upwards (downwards)
renormalization of $J$ ($J'$) compared to LSWT.) The dependence of the
magnon energy on the $c$-component of the wavevector is weak, giving
the dispersion a pronounced one-dimensional character.

\subsubsection{Comparison with experimental dispersion}

In Fig. \ref{fig-th-exp} we also show experimental results for the
magnon energies obtained from inelastic neutron
scattering.\cite{coldea} The open symbols are data points
corresponding to the positions of the \textit{strongest} peaks in
the unpolarized neutron scattering data. The red dashed line is a
fit of these points to a dispersion given by
\begin{equation}
\omega_{\bm{k}}=\sqrt{(\tilde{J}_{\bm{k}}-\tilde{J}_{\bm{Q}})\Big[(\tilde{J}_{\bm{Q}+\bm{k}}+\tilde{J}_{\bm{Q}-\bm{k}})
/2-\tilde{J}_{\bm{Q}}\Big]}, \label{lswtdisp}
\end{equation}
where $\tilde{J}_{\bm{k}}=\tilde{J}\cos k_b + 2\tilde{J'}\cos
(k_b/2)\cos(\sqrt{3}k_c/2)$. The functional form of Eq.
(\ref{lswtdisp}) is identical to the expression from linear
spin-wave (LSWT) theory in the absence of the DM interaction, but in
Eq. (\ref{lswtdisp}) the bare exchange parameters $J$ and $J'$ that
would be used in LSWT have been replaced by \textit{effective},
renormalized parameters ($\tilde{J}$ and $\tilde{J}'$, respectively)
whose values are chosen to give the best fit to the experimental
data points. Hence the ratios of these renormalized values to the
bare values give a measure of the strength of the quantum
fluctuations in the system. In Fig. \ref{fig-th-exp}
$\tilde{J}=0.61(1)$ meV and $\tilde{J}'=0.107(10)$ meV.\cite{coldea}

\begin{figure}[!htb]
\begin{center}
  \includegraphics[width=8cm]{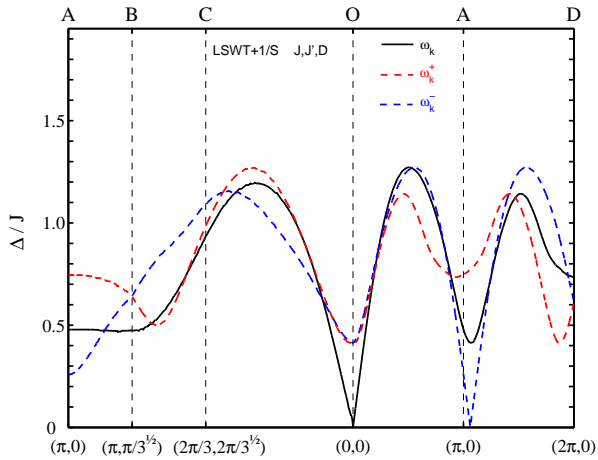}
  \caption{\label{3modes}
  (Color online) The magnon modes $\omega_{\bm{k}}$ (principal mode) and
  $\omega_{\bm{k}\pm\bm{Q}}\equiv \omega_{\bm{k}}^{\pm}$
  (secondary modes), as calculated from spin-wave theory
  with $1/S$ corrections,\cite{veillette} for Cs$_2$CuCl$_4$ ($J/J'=2.92$ and
  $D/J'=0.16$), plotted along the path ABCOAD shown in Fig. \ref{trlattbz}.}
\end{center}
\end{figure}

Before comparing the theoretical and experimental curves, however,
we first discuss the interpretation of the experimental data in some
more detail, i.e. why for a given $\bm{k}$ the location in energy of
the strongest peak in the data can in most cases be identified with
the magnon energy $\omega_{\bm{k}}$. To this end, we consider the
differential cross section for inelastic scattering of unpolarized
neutrons, which is proportional to\cite{lovesey}
\be
\sin^2 \theta_{\bm{k}}\,S^{aa}(\bm{k},\omega)+(1+\cos^2
\theta_{\bm{k}})\,S^{bb}(\bm{k},\omega)
\label{cs1}
\ee
where $\theta_{\bm{k}}$ is the angle between the scattering wave vector
$\bm{k}$ and the axis ($\hat{a}$) perpendicular to the ordering
($bc$) plane of the spins, and $S^{aa}(\bm{k},\omega)$ and
$S^{bb}(\bm{k},\omega)$ are diagonal components of the dynamical
structure factor defined as ($\mu,\nu=a,b,c$)
\be
S^{\mu\nu}(\bm{k},\omega) =
\frac{1}{2\pi}\int_{-\infty}^{\infty}dt\;\sum_{\bm{R}}\langle
S^{\mu}(0,0)S^{\nu}(\bm{R},t) \rangle e^{-i(\bm{k}\cdot\bm{R}-\omega
t)}.
\ee
When $\bm{k}$ is in the $bc$ plane (as in our
calculations), $\theta_{\bm{k}}=\pi/2$, in which case Eq.
(\ref{cs1}) reduces to
\be
S^{aa}(\bm{k},\omega)+S^{bb}(\bm{k},\omega).
\label{cs2}
\ee
Spin-wave theory predicts sharp one-magnon peaks in the out-of-plane
correlations $S^{aa}(\bm{k},\omega)$ at $\omega=\omega_{\bm{k}}$
(referred to as the principal mode) and in the in-plane correlations
$S^{bb}(\bm{k},\omega)$ at both $\omega=\omega_{\bm{k}+\bm{Q}}
\equiv \omega_{\bm{k}}^+$ and $\omega=\omega_{\bm{k}-\bm{Q}}\equiv
\omega_{\bm{k}}^-$ (referred to as the two secondary modes), where
$\omega_{\bm{k}}$ is the magnon dispersion. Hence if the spin-wave
prediction for Eq. (\ref{cs2}) were plotted as a function of
$\omega$ for fixed $\bm{k}$, three peaks would be observed, at
$\omega=\omega_{\bm{k}}$ and $\omega_{\bm{k}}^{\pm}$. Although
spin-wave theory does not give the correct $\bm{k}$-dependence of
the three modes, it is expected that the three-peak structure it
predicts (for a system with helical magnetic order) is qualitatively
correct. In Fig. \ref{3modes} we have plotted this structure for
$\bm{k}$ varying along the path in Fig. \ref{trlattbz} (the energy
of the modes was calculated from spin-wave theory with $1/S$
corrections\cite{veillette}). For example, at point B the spin-wave calculation
predicts that $\omega_{\bm{k}}$ lies well below the two other modes
in energy. This principal mode is also predicted to have the
strongest intensity at B. Thus it appears quite safe to assume that
the lowest-energy, strongest-intensity peak in the experimental
spectrum at that $\bm{k}$-vector should be assigned to
$\omega_{\bm{k}}$. At high energies the intensity of the
out-of-plane correlations $S^{aa}$ is also predicted to be bigger
which justifies an assignment of the location of the strongest peak
to $\omega_{\bm{k}}$ at high energies. On the other hand, at low
energies close to $\bm{k}=\bm{Q}$, where the primary mode
$\omega_{\bm{k}}$ is expected to be gapped and located inside the
V-shape of the $\omega_{\bm{k}}^-$ mode (see Fig. \ref{3modes}), the
$S^{aa}$ part can not be separately resolved. The experimental data
indicate that there is no (or very little) overall gap at $k=Q$.
This is what one would expect, because for a spiral with finite DM
interaction one of the in-plane modes ($\omega_{\bm{k}}^{-}$) would
still be gapless.

\begin{figure}[!htb]
\begin{center}
\includegraphics[width=8cm]{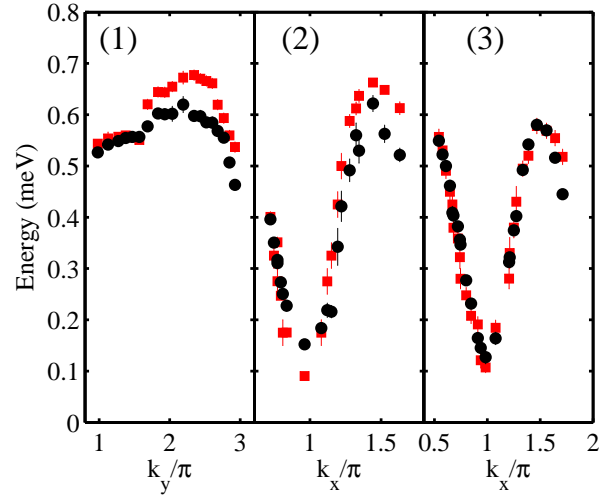}
  \caption{\label{further_cs2cucl4_comparison}
(Color online) Comparison of the experimentally measured
dispersion relation in Cs$_2$CuCl$_4$ (red squares) and series expansion results for
$J=0.385$ meV, $J/J'=2.92$ and $D/J'=0.16$ (black circles), along paths (1), (2) and
(3) in Fig. \ref{trlattbz}. Calculations were
made at the same ($k_x,k_y$) where experimental dispersion points
were measured.}
\end{center}
\end{figure}

For point B, $\omega_{\bm{k}}$ is quite unambiguously determined, as
discussed above, and this point is sensitive to the presence of the
DM interaction; the theoretically calculated energy increases with
$D$. This makes the interchain dispersion in better agreement with
the experimental data when $D$ is included, although there is still
a slight overestimate of $\sim 20\%$. Along the AD direction the
agreement between experiments and series is essentially perfect,
while along the chain direction OA, where the dispersion is maximal,
the theory makes a slight underestimate of $\sim 10\%$. The
experimental dispersion relation was also measured along
off-symmetry directions in the 2D Brillouin zone and Fig.\
\ref{further_cs2cucl4_comparison} shows that the series results
(black circles) compare very well with experimental data (red
squares) for those directions too.

\begin{figure}[!htb]
\begin{center}
  \includegraphics[width=7cm,bbllx=20,bblly=151,bburx=553,
  bbury=687,angle=0,clip=]{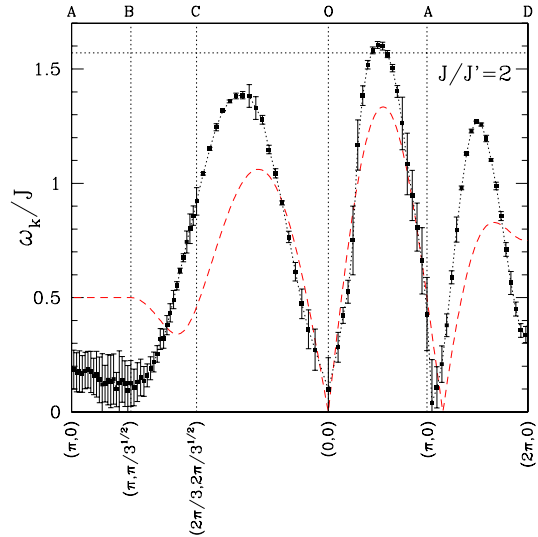}
  \caption{\label{fig_mk_y2}
(Color online) Magnon dispersion for Cs$_2$CuBr$_4$ ($J=2J'$ and
$D=0$), calculated from series expansions. Compared to the
dispersion from linear spin-wave theory (red dashed line, Eq.
(\ref{lswtdisp})) the series dispersion is enhanced along the $J$
bonds and decreased perpendicular to them. The dispersion looks
qualitatively rather similar to the case $J/J'=2.92$, $D=0$ in Fig.
\ref{fig_mk_y3} although in that case the energy is practically flat
along AB (i.e., perpendicular to the chains) while here the energy
varies somewhat along AB, giving a less one-dimensional dispersion,
as expected. The horizontal black dashed line is at $\omega/J=\pi/2$
which is the maximum energy in the one-dimensional limit
$J/J'=\infty$.}
\end{center}
\end{figure}

\subsection{Cs$_2$CuBr$_4$}

In Fig. \ref{fig_mk_y2} we show series expansion results (black
symbols) for the magnon dispersion for $J/J'=2$, the ratio
appropriate for Cs$_2$CuBr$_4$, along the $\bm{k}$-space path ABCOAD
in Fig. \ref{trlattbz}. The dispersion obtained from linear
spin-wave theory, Eq. (\ref{lswtdisp}) is also shown for comparison.
We note the following features: (i) With respect to LSWT the series
dispersion is renormalized \textit{upwards} in the direction
parallel to the chains (i.e., in the direction along which the
exchange constant is strongest) and \textit{downwards} in the
perpendicular direction. A similar kind of upward renormalization
occurs both for the square lattice ($J=0$) and decoupled chains
($J'=0$). (ii) Along the AB direction perpendicular to the chains,
the series dispersion has flattened and the excitation energy has
decreased significantly compared to the case $J=J'$ case (Fig. 1 in
Ref. \onlinecite{zheng06}). Features (i) and (ii) persist as $J/J'$
is increased further to 3, as seen in the $D=0$ curve in Fig.
\ref{fig_mk_y3}, which except for the more pronounced flatness along
AB looks qualitatively very similar to the $J/J'=2$ dispersion in Fig.
\ref{fig_mk_y2}.

\section{Conclusions}
\label{concl}

In this paper we have presented series expansion calculations for
various ground state properties (ground state energy, ordering
wavevector, and sublattice magnetization), and for the magnon
dispersion, in the helically ordered state of spin-$1/2$ frustrated
antiferromagnets on an anisotropic triangular lattice. For the
parameter values considered here the model Hamiltonians we have
studied are expected to give a good description of the magnetic
properties of Cs$_2$CuBr$_4$ and Cs$_2$CuCl$_4$.

For parameters appropriate for Cs$_2$CuBr$_4$ ($J/J' = 2$) we have
made specific predictions for large and non-uniform renormalizations
of the dispersion relation compared to the classical (linear spin-wave
theory) result (see Fig.\ \ref{fig_mk_y2}). There is an upward renormalization
along the chains and a downward renormalization along the interchain
direction. These predictions for Cs$_2$CuBr$_4$ should be testable by
future neutron scattering experiments.

For Cs$_2$CuCl$_4$ we have compared our theoretical predictions for
the dispersion with experimental results from neutron scattering
experiments and with spin-wave theory that includes $1/S$ corrections.
The calculated magnon dispersion shows large quantum renormalizations
compared to the classical result; as for Cs$_2$CuBr$_4$ the renormalization
is upward along the chain direction and downward along the interchain
direction. Thus quantum fluctuations make the dispersion look more
one-dimensional. These predictions from series are in good
quantitative agreement with the experimental observations. In
contrast, spin-wave theory with $1/S$ corrections predicts
considerably smaller renormalizations and thus underestimates the
effects of quantum fluctuations on the dispersion.

The agreement between theory and experiment improves further when
the Dzyaloshinskii-Moriya (DM) interaction is taken into account.
The DM interaction is found to significantly increase the
incommensuration of the ordering wavevector with respect to the
antiferromagnetic wavevector in the chain ($\hat{b}$) direction.
Another important consequence of the DM interaction is that it opens
up a gap in the magnon dispersion at the ordering wavevector. We
have made specific predictions for the size of this energy gap at
the ordering wavevector to longitudinally-polarized excitations
(along the direction normal to the spiral plane) which could be
tested by future high-resolution {\em polarized} neutron scattering
experiments.

Another very interesting issue, at least from a theoretical point
of view, is the role of the DM interaction in Cs$_2$CuCl$_4$
in stabilizing a magnetically ordered ground state. We find here
that inclusion of the DM interaction strengthens the magnetic ordering.
In fact, several recent papers\cite{dalidovich,yunsor,alicea,sb} have
predicted that in the absence of the DM interaction the ground state
is not magnetically ordered. The proposed alternatives
to a magnetically ordered phase for $D=0$ include a spin liquid
(``algebraic vortex liquid'')\cite{alicea} and a state with
``staggered'' dimerization.\cite{sb} The latter state could be
studied using series expansions; it would be interesting to compare
its energy with the helically ordered phase. The fact that a
different dimerized state (with ``diagonal'' dimerization; see Fig.
4a in Ref. \onlinecite{zheng99}) that we have studied with series
expansions is extremely close in energy to the helically ordered
state (for a large regime of $J/J'$ including $J/J'\approx 3$ we
find that the energies of the two phases lie virtually on top of
each other) is strongly suggestive that the state with staggered
dimerization proposed in Ref. \onlinecite{sb} may be highly
competitive as well.

\begin{acknowledgments}

We thank G. Aeppli, F. Becca, S. Hayden, D. McMorrow, B. Powell, S.
Sorella, and M. Veillette for helpful discussions. This work was
supported by the Australian Research Council (ZW, JOF, and RHM), the
US National Science Foundation, Grant No. DMR-0240918 (RRPS), and
the United Kingdom Engineering and Physical Sciences Research
Council, Grant No. GR/R76714/02 (RC). RHM thanks UC Davis for
hospitality. We are grateful for the computing resources provided by
the Australian Partnership for Advanced Computing (APAC) National
Facility and by the Australian Centre for Advanced Computing and
Communications (AC3).

\end{acknowledgments}

\widetext

\begin{table}
\squeezetable \caption{Series coefficients for the magnon dispersion
for the isotropic triangular-lattice model, calculated for $J=2.92$,
$J'=1$, $q=1.64$, $D=0.16$, and $t=4$ in Eqs. (\ref{H0}) and
(\ref{V}). Nonzero coefficients $c_{r,m,n}$ in Eq. (\ref{eq_mk}) up
to order $r=8$ are listed.} \label{tab_mk}
\begin{tabular}{|ll|ll|ll|ll|}
\hline \hline \multicolumn{1}{|c}{($r,m,n$)}
&\multicolumn{1}{c|}{$c_{r,m,n}$} &\multicolumn{1}{c}{($r,m,n$)}
&\multicolumn{1}{c|}{$c_{r,m,n}$} &\multicolumn{1}{c}{($r,m,n$)}
&\multicolumn{1}{c|}{$c_{r,m,n}$}
&\multicolumn{1}{c}{($r,m,n$)} &\multicolumn{1}{c|}{$c_{r,m,n}$} \\
\hline
 ( 0, 0, 0) &   7.349606933       &( 4, 1, 3) &   7.578150248$\times 10^{-4}$ &( 4, 7, 1) &  -9.342152056$\times 10^{-3}$ &( 6, 7, 5) &  -1.670797799$\times 10^{-6}$  \\
 ( 1, 0, 0) &  -4.000000000       &( 5, 1, 3) &  -9.976179492$\times 10^{-4}$ &( 5, 7, 1) &  -1.835064338$\times 10^{-2}$ &( 7, 7, 5) &   1.015387605$\times 10^{-6}$  \\
 ( 2, 0, 0) &   6.038563795$\times 10^{-2}$ &( 6, 1, 3) &  -2.926714357$\times 10^{-3}$ &( 6, 7, 1) &  -2.367732439$\times 10^{-2}$ &( 8, 7, 5) &   5.830972082$\times 10^{-6}$  \\
 ( 3, 0, 0) &  -6.217604565$\times 10^{-2}$ &( 7, 1, 3) &  -4.924224688$\times 10^{-3}$ &( 7, 7, 1) &  -2.521951807$\times 10^{-2}$ &( 6, 8, 4) &  -1.600507150$\times 10^{-5}$  \\
 ( 4, 0, 0) &  -1.125358729$\times 10^{-1}$ &( 8, 1, 3) &  -6.508161119$\times 10^{-3}$ &( 8, 7, 1) &  -2.415728150$\times 10^{-2}$ &( 7, 8, 4) &  -2.409589103$\times 10^{-5}$  \\
 ( 5, 0, 0) &  -1.162865812$\times 10^{-1}$ &( 3, 3, 3) &   6.950474681$\times 10^{-4}$ &( 4, 8, 0) &  -6.795995712$\times 10^{-3}$ &( 8, 8, 4) &  -2.189863027$\times 10^{-5}$  \\
 ( 6, 0, 0) &  -9.649286838$\times 10^{-2}$ &( 4, 3, 3) &  -9.066263964$\times 10^{-4}$ &( 5, 8, 0) &  -1.353032723$\times 10^{-2}$ &( 6, 9, 3) &  -8.269690490$\times 10^{-5}$  \\
 ( 7, 0, 0) &  -6.982209428$\times 10^{-2}$ &( 5, 3, 3) &  -1.496710785$\times 10^{-3}$ &( 6, 8, 0) &  -1.705500588$\times 10^{-2}$ &( 7, 9, 3) &  -1.831650139$\times 10^{-4}$  \\
 ( 8, 0, 0) &  -4.568446552$\times 10^{-2}$ &( 6, 3, 3) &  -5.060095359$\times 10^{-4}$ &( 7, 8, 0) &  -1.715246720$\times 10^{-2}$ &( 8, 9, 3) &  -2.277789431$\times 10^{-4}$  \\
 ( 1, 1, 1) &   7.712345303$\times 10^{-1}$ &( 7, 3, 3) &   1.528045822$\times 10^{-3}$ &( 8, 8, 0) &  -1.507009210$\times 10^{-2}$ &( 6,10, 2) &  -2.517791809$\times 10^{-4}$  \\
 ( 2, 1, 1) &  -1.685921734$\times 10^{-1}$ &( 8, 3, 3) &   3.781329536$\times 10^{-3}$ &( 5, 1, 5) &   1.418417481$\times 10^{-5}$ &( 7,10, 2) &  -6.579095073$\times 10^{-4}$  \\
 ( 3, 1, 1) &  -2.106120525$\times 10^{-1}$ &( 3, 4, 2) &   8.064630409$\times 10^{-3}$ &( 6, 1, 5) &   7.457692880$\times 10^{-6}$ &( 8,10, 2) &  -9.157527309$\times 10^{-4}$  \\
 ( 4, 1, 1) &  -1.865808906$\times 10^{-1}$ &( 4, 4, 2) &   8.310784630$\times 10^{-3}$ &( 7, 1, 5) &  -6.287690575$\times 10^{-6}$ &( 6,11, 1) &  -4.658726892$\times 10^{-4}$  \\
 ( 5, 1, 1) &  -1.379395553$\times 10^{-1}$ &( 5, 4, 2) &   9.603610069$\times 10^{-3}$ &( 8, 1, 5) &  -1.316193701$\times 10^{-5}$ &( 7,11, 1) &  -1.474734326$\times 10^{-3}$  \\
 ( 6, 1, 1) &  -9.056954635$\times 10^{-2}$ &( 6, 4, 2) &   1.286421148$\times 10^{-2}$ &( 5, 3, 5) &   7.092087407$\times 10^{-6}$ &( 8,11, 1) &  -2.737063946$\times 10^{-3}$  \\
 ( 7, 1, 1) &  -5.546855043$\times 10^{-2}$ &( 7, 4, 2) &   1.664667471$\times 10^{-2}$ &( 6, 3, 5) &  -2.077134408$\times 10^{-6}$ &( 6,12, 0) &  -2.330119692$\times 10^{-4}$  \\
 ( 8, 1, 1) &  -3.373111871$\times 10^{-2}$ &( 8, 4, 2) &   1.954078358$\times 10^{-2}$ &( 7, 3, 5) &  -1.309394434$\times 10^{-5}$ &( 7,12, 0) &  -8.136447772$\times 10^{-4}$  \\
 ( 1, 2, 0) &   1.396200322$\times 10^{-2}$ &( 3, 5, 1) &   2.242175366$\times 10^{-2}$ &( 8, 3, 5) &  -1.292053918$\times 10^{-5}$ &( 8,12, 0) &  -1.712422519$\times 10^{-3}$  \\
 ( 2, 2, 0) &  -1.505333368$\times 10^{-2}$ &( 4, 5, 1) &   1.427117691$\times 10^{-2}$ &( 5, 5, 5) &   1.418417481$\times 10^{-6}$ &( 7, 1, 7) &   1.364242160$\times 10^{-7}$  \\
 ( 3, 2, 0) &  -2.815674423$\times 10^{-4}$ &( 5, 5, 1) &  -2.602862242$\times 10^{-3}$ &( 6, 5, 5) &  -5.076936139$\times 10^{-6}$ &( 8, 1, 7) &   1.106865429$\times 10^{-7}$  \\
 ( 4, 2, 0) &   1.594690655$\times 10^{-2}$ &( 6, 5, 1) &  -1.521289949$\times 10^{-2}$ &( 7, 5, 5) &  -7.060086129$\times 10^{-6}$ &( 7, 3, 7) &   8.185452960$\times 10^{-8}$  \\
 ( 5, 2, 0) &   2.998362910$\times 10^{-2}$ &( 7, 5, 1) &  -1.993026012$\times 10^{-2}$ &( 8, 5, 5) &   1.083485410$\times 10^{-6}$ &( 8, 3, 7) &   1.069367604$\times 10^{-8}$  \\
 ( 6, 2, 0) &   3.941680852$\times 10^{-2}$ &( 8, 5, 1) &  -1.888733912$\times 10^{-2}$ &( 5, 6, 4) &   2.746050501$\times 10^{-5}$ &( 7, 5, 7) &   2.728484320$\times 10^{-8}$  \\
 ( 7, 2, 0) &   4.308338742$\times 10^{-2}$ &( 3, 6, 0) &   4.006413979$\times 10^{-4}$ &( 6, 6, 4) &   1.039487431$\times 10^{-5}$ &( 8, 5, 7) &  -4.924369043$\times 10^{-8}$  \\
 ( 8, 2, 0) &   4.120905564$\times 10^{-2}$ &( 4, 6, 0) &  -3.331359320$\times 10^{-3}$ &( 7, 6, 4) &  -1.115844752$\times 10^{-5}$ &( 7, 7, 7) &   3.897834743$\times 10^{-9}$  \\
 ( 2, 0, 2) &  -2.650432163$\times 10^{-2}$ &( 5, 6, 0) &  -6.163747248$\times 10^{-3}$ &( 8, 6, 4) &   6.482844962$\times 10^{-6}$ &( 8, 7, 7) &  -3.233550435$\times 10^{-8}$  \\
 ( 3, 0, 2) &   3.538137283$\times 10^{-3}$ &( 6, 6, 0) &  -7.199966978$\times 10^{-3}$ &( 5, 7, 3) &   2.089666851$\times 10^{-4}$ &( 7, 8, 6) &   1.034023057$\times 10^{-7}$  \\
 ( 4, 0, 2) &   2.465629073$\times 10^{-3}$ &( 7, 6, 0) &  -7.409354256$\times 10^{-3}$ &( 6, 7, 3) &   2.467736333$\times 10^{-4}$ &( 8, 8, 6) &  -1.557033299$\times 10^{-7}$  \\
 ( 5, 0, 2) &  -1.050320907$\times 10^{-2}$ &( 8, 6, 0) &  -8.061924799$\times 10^{-3}$ &( 7, 7, 3) &  -3.668402903$\times 10^{-5}$ &( 7, 9, 5) &   1.173239911$\times 10^{-6}$  \\
 ( 6, 0, 2) &  -2.242983874$\times 10^{-2}$ &( 4, 0, 4) &  -9.054115944$\times 10^{-5}$ &( 8, 7, 3) &  -5.809689111$\times 10^{-4}$ &( 8, 9, 5) &   7.836228877$\times 10^{-7}$  \\
 ( 7, 0, 2) &  -2.768399107$\times 10^{-2}$ &( 5, 0, 4) &  -5.479146209$\times 10^{-5}$ &( 5, 8, 2) &   7.675228288$\times 10^{-4}$ &( 7,10, 4) &   7.377737697$\times 10^{-6}$  \\
 ( 8, 0, 2) &  -2.619803879$\times 10^{-2}$ &( 6, 0, 4) &   3.687786670$\times 10^{-6}$ &( 6, 8, 2) &   1.512127062$\times 10^{-3}$ &( 8,10, 4) &   1.539074513$\times 10^{-5}$  \\
 ( 2, 2, 2) &  -2.650432163$\times 10^{-2}$ &( 7, 0, 4) &  -1.852548500$\times 10^{-5}$ &( 7, 8, 2) &   1.521619340$\times 10^{-3}$ &( 7,11, 3) &   2.754833548$\times 10^{-5}$  \\
 ( 3, 2, 2) &  -9.979720263$\times 10^{-4}$ &( 8, 0, 4) &  -1.568789895$\times 10^{-4}$ &( 8, 8, 2) &   6.001765269$\times 10^{-4}$ &( 8,11, 3) &   8.079901649$\times 10^{-5}$  \\
 ( 4, 2, 2) &   6.636359222$\times 10^{-3}$ &( 4, 2, 4) &  -1.207215459$\times 10^{-4}$ &( 5, 9, 1) &   1.165606703$\times 10^{-3}$ &( 7,12, 2) &   5.930561117$\times 10^{-5}$  \\
 ( 5, 2, 2) &   5.391428994$\times 10^{-3}$ &( 5, 2, 4) &  -2.149262987$\times 10^{-5}$ &( 6, 9, 1) &   2.854089455$\times 10^{-3}$ &( 8,12, 2) &   1.954393847$\times 10^{-4}$  \\
 ( 6, 2, 2) &   2.036622085$\times 10^{-3}$ &( 6, 2, 4) &   7.358270843$\times 10^{-5}$ &( 7, 9, 1) &   4.479815788$\times 10^{-3}$ &( 7,13, 1) &   5.956579054$\times 10^{-5}$  \\
 ( 7, 2, 2) &   1.434767818$\times 10^{-4}$ &( 7, 2, 4) &   5.557974820$\times 10^{-5}$ &( 8, 9, 1) &   5.794251007$\times 10^{-3}$ &( 8,13, 1) &   1.916748738$\times 10^{-4}$  \\
 ( 8, 2, 2) &   6.693067703$\times 10^{-4}$ &( 8, 2, 4) &  -3.425847619$\times 10^{-5}$ &( 5,10, 0) &   2.060334673$\times 10^{-5}$ &( 7,14, 0) &   1.040120579$\times 10^{-6}$  \\
 ( 2, 3, 1) &  -3.084059676$\times 10^{-1}$ &( 4, 4, 4) &  -3.018038648$\times 10^{-5}$ &( 6,10, 0) &  -1.952180586$\times 10^{-4}$ &( 8,14, 0) &  -1.495168851$\times 10^{-5}$  \\
 ( 3, 3, 1) &  -1.588761190$\times 10^{-1}$ &( 5, 4, 4) &   6.075933723$\times 10^{-5}$ &( 7,10, 0) &  -6.035252148$\times 10^{-4}$ &( 8, 0, 8) &  -7.586845473$\times 10^{-9}$  \\
 ( 4, 3, 1) &  -4.101867358$\times 10^{-2}$ &( 6, 4, 4) &   1.000503366$\times 10^{-4}$ &( 8,10, 0) &  -9.362009169$\times 10^{-4}$ &( 8, 2, 8) &  -1.213894466$\times 10^{-8}$  \\
 ( 5, 3, 1) &   3.354266372$\times 10^{-2}$ &( 7, 4, 4) &   9.862298579$\times 10^{-5}$ &( 6, 0, 6) &  -7.263598281$\times 10^{-7}$ &( 8, 4, 8) &  -6.069460194$\times 10^{-9}$  \\
 ( 6, 3, 1) &   7.243569855$\times 10^{-2}$ &( 8, 4, 4) &   1.673415457$\times 10^{-4}$ &( 7, 0, 6) &  -5.650646253$\times 10^{-7}$ &( 8, 6, 8) &  -1.734128016$\times 10^{-9}$  \\
 ( 7, 3, 1) &   8.854496129$\times 10^{-2}$ &( 4, 5, 3) &  -4.901068912$\times 10^{-4}$ &( 8, 0, 6) &   2.353281314$\times 10^{-7}$ &( 8, 8, 8) &  -2.167670129$\times 10^{-10}$  \\
 ( 8, 3, 1) &   9.234591989$\times 10^{-2}$ &( 5, 5, 3) &   9.426485008$\times 10^{-5}$ &( 6, 2, 6) &  -1.089539742$\times 10^{-6}$ &( 8, 9, 7) &  -6.526162920$\times 10^{-9}$  \\
 ( 2, 4, 0) &  -4.736398330$\times 10^{-1}$ &( 6, 5, 3) &   1.397521826$\times 10^{-3}$ &( 7, 2, 6) &  -4.743845804$\times 10^{-7}$ &( 8,10, 6) &  -8.596029431$\times 10^{-8}$  \\
 ( 3, 4, 0) &  -4.199253740$\times 10^{-1}$ &( 7, 5, 3) &   2.688369202$\times 10^{-3}$ &( 8, 2, 6) &   1.060978601$\times 10^{-6}$ &( 8,11, 5) &  -6.482179010$\times 10^{-7}$  \\
 ( 4, 4, 0) &  -3.246293814$\times 10^{-1}$ &( 8, 5, 3) &   3.441529158$\times 10^{-3}$ &( 6, 4, 6) &  -4.358158969$\times 10^{-7}$ &( 8,12, 4) &  -3.069805782$\times 10^{-6}$  \\
 ( 5, 4, 0) &  -2.188669309$\times 10^{-1}$ &( 4, 6, 2) &  -3.058604819$\times 10^{-3}$ &( 7, 4, 6) &   4.028642250$\times 10^{-7}$ &( 8,13, 3) &  -9.416441458$\times 10^{-6}$  \\
 ( 6, 4, 0) &  -1.251987959$\times 10^{-1}$ &( 5, 6, 2) &  -4.021118013$\times 10^{-3}$ &( 8, 4, 6) &   1.336492139$\times 10^{-6}$ &( 8,14, 2) &  -1.872276535$\times 10^{-5}$  \\
 ( 7, 4, 0) &  -5.434016523$\times 10^{-2}$ &( 6, 6, 2) &  -2.960843639$\times 10^{-3}$ &( 6, 6, 6) &  -7.263598281$\times 10^{-8}$ &( 8,15, 1) &  -2.388666059$\times 10^{-5}$  \\
 ( 8, 4, 0) &  -7.501154228$\times 10^{-3}$ &( 7, 6, 2) &  -1.581263941$\times 10^{-3}$ &( 7, 6, 6) &   4.155864858$\times 10^{-7}$ &( 8,16, 0) &  -8.823605179$\times 10^{-6}$  \\
 ( 3, 1, 3) &   2.085142404$\times 10^{-3}$ &( 8, 6, 2) &  -1.344189910$\times 10^{-3}$ &( 8, 6, 6) &   4.410976005$\times 10^{-7}$         \\
\hline \hline
\end{tabular}
\end{table}

\end{document}